\documentclass[runningheads]{llncs}
\usepackage[T1]{fontenc}
\usepackage{graphicx}
\usepackage{tikz}
\usetikzlibrary{shapes.geometric, arrows}
\usepackage{multirow}
\usepackage{gensymb}
\usepackage{wasysym}
\usepackage{subcaption}
\usepackage[normalem]{ulem}
\usepackage[textsize=tiny]{todonotes}

\begin{document}
\title{Application of Graph Neural Networks in \\Dark Photon Search with Visible Decays at Future Beam Dump Experiment} 
%
\titlerunning{GNN Application in Dark Photon Search at Future Beam Dump Experiment}

\author{Zejia Lu\inst{1,2,3} \and
Xiang Chen\inst{1,2,3} \and
Jiahui Wu\inst{1,2,3} \and
Yulei Zhang\inst{1,2,3}\thanks{Corresponding author. avencast@sjtu.edu.cn.} \and
Liang Li\inst{1,2,3}\thanks{Corresponding author. liangliphy@sjtu.edu.cn.}
}

\authorrunning{Z.~Lu et. al}
%
\institute{School of Physics and Astronomy, Institute of Nuclear and Particle Physics,
Shanghai Jiao Tong University, Shanghai, China
\and
Shanghai Key Laboratory for Particle Physics and Cosmology
\and
Key Lab for Particle Physics,
Astrophysics and Cosmology (MOE) \\
}
\maketitle              
\begin{abstract}
Beam dump experiments provide a distinctive opportunity to search for dark photons, which are compelling candidates for dark matter with low mass. In this study, we propose the application of Graph Neural Networks (GNN) in tracking reconstruction with beam dump experiments to obtain high resolution in both tracking and vertex reconstruction. Our findings demonstrate that 
in a typical 3-track scenario with the visible decay mode, the GNN approach significantly outperforms the traditional approach, improving the 3-track reconstruction efficiency by up to 88\% in the low mass region. 
Furthermore, we show that improving the minimal vertex detection distance significantly impacts the signal sensitivity in dark photon searches with the visible decay mode. 
By reducing the minimal vertex distance from 5 mm to 0.1 mm, the exclusion upper limit on the dark photon mass ($m_A\prime$) can be improved by up to a factor of 3.

\end{abstract}

\section{Introduction}
Dark matter is one of the most intriguing mysterious that cannot be solved by the Standard Model. Numerous astrophysical and cosmological observations have provided strong evidence for the presence of dark matter, which is believed to make up about 85\% of the matter in the universe. 
Despite its prevalence, the nature of dark matter particles remains elusive, and identifying their properties is a key challenge in particle physics research. 
Direct searches for Weakly Interacting Massive Particles (WIMPs) as one of most sought-after dark matter candidates have so far yielded null results~\cite{pandax}. 
This outcome has further motivated the exploration of alternative dark matter candidates, such as the dark photon, a hypothetical particle also known as a U(1) gauge boson of a hidden sector and is predicted to have a small mass and interact weakly with standard matter.  
Unlike WIMPs, the dark photon is predicted to have extremely weak interactions with standard matter, making it a challenging particle to detect using traditional experimental methods. 
Numerous experimental efforts are underway to search for dark photons in various decay channels and interaction modes~\cite{ldmx_exp,NA64,darkshine_exp}.

\begin{figure}[htbp]
  \centering
  \begin{subfigure}{0.45\textwidth}
    \includegraphics[width=\linewidth]{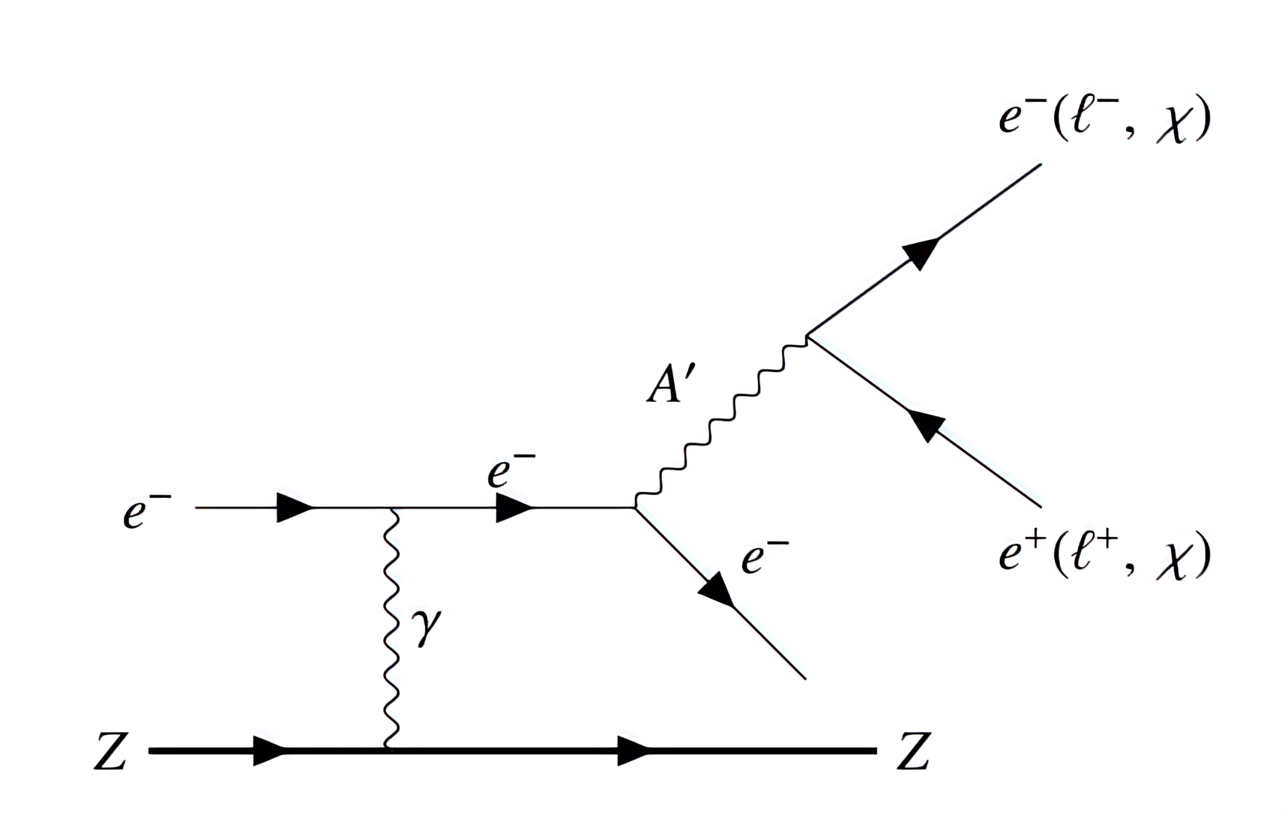}
    \caption{}
    \label{fig:feynman_signal}
  \end{subfigure}
  \hfill
  \begin{subfigure}{0.45\textwidth}
    \includegraphics[width=\linewidth]{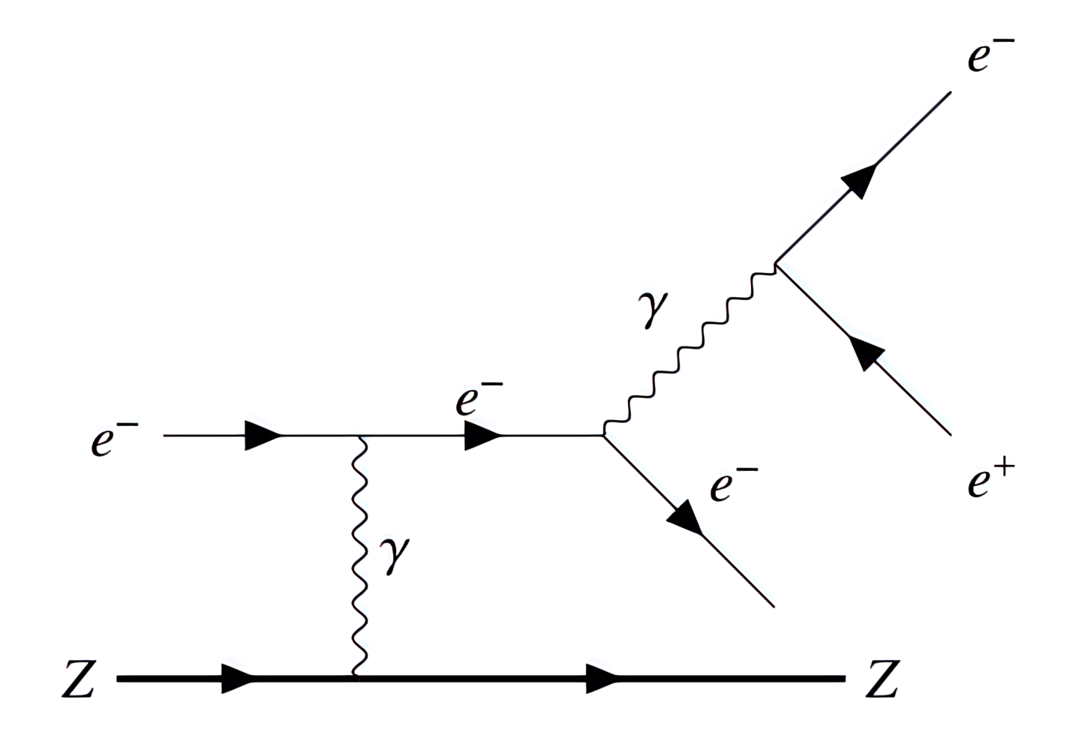}
    \caption{}
    \label{fig:feynman_bkg}
  \end{subfigure}
  \caption{Feynman diagram of (a) dark photon produced through bremsstrahlung and decaying into leptons pair or dark matter $\chi$ (b) QED radiative trident reaction, which serves as main background of dark photon visible decay~\cite{searching}. $A'$ is dark photon while $\gamma$ is photon.}
  \label{fig:points}
\end{figure}

Beam dump experiments provide a unique opportunity to explore dark photons as dark matter candidates due to their distinctive experimental setup. In a beam dump experiment, a high-energy particle beam is directed towards a target, where interactions between the beam particles and target nuclei can produce new particles, including dark photons, as illustrated in Fig.~\ref{fig:feynman_signal}.
The generated dark photons can escape the target and travel through the experimental setup before decaying into visible particles that can be detected by the surrounding detectors.
One of the key advantages of beam dump experiments is their ability to probe a wide range of dark photon masses and couplings. 
Unlike traditional direct detection experiments that are sensitive to a specific mass range, beam dump experiments have the potential to cover a broad spectrum of dark photon masses, including those that are challenging to access with other experimental methods.
Furthermore, beam dump experiments are relatively cost-effective and can be conducted using existing accelerator facilities, making them attractive platforms for exploring new physics beyond the Standard Model. 

Dark photons couple with standard matter through the kinetic mixing term~\cite{searching}
\begin{equation}
    \mathcal{L}_{dark,\gamma}=-e\epsilon A'_{\mu} J^\mu_{em},
\end{equation}
where $\epsilon$ is the coupling constant, $A'_{\mu}$ is the mediator field of the dark U(1) gauge group. 
The model has two free parameters, the coupling constant $\epsilon$ and the dark photon mass $m_{A'}$. 

By comparing measured decay products with expected backgrounds, beam dump experiments can provide crucial insights into the nature of dark photons, 
either supporting their existence within specific parameter regions or setting stringent limits on these parameters.
This process involves dark photon signature searches, background estimation, sensitivity projections, and statistical analyses. 
With advanced detector technologies and innovative analysis techniques like Graph Neural Networks, the sensitivity to dark photon signals can be significantly improved, further 
motivating the search for these elusive particles in beam dump experiments.

\section{Experimental Setup and Simulation Framework}
 Several beam-dumped experiments have been proposed to search for the dark photon, such as LDMX\cite{ldmx_exp}, NA64\cite{NA64}, and DarkSHINE experiments\cite{darkshine_exp}.
 The experimental setup of our study is mainly based on the DarkSHINE experiment, an electron-on-fixed-target experiment using a single electron beam. 
 
 \subsection{Experimental Setup}
 
\begin{figure}[!htb]
    \centering
    \includegraphics[width=0.8\textwidth]{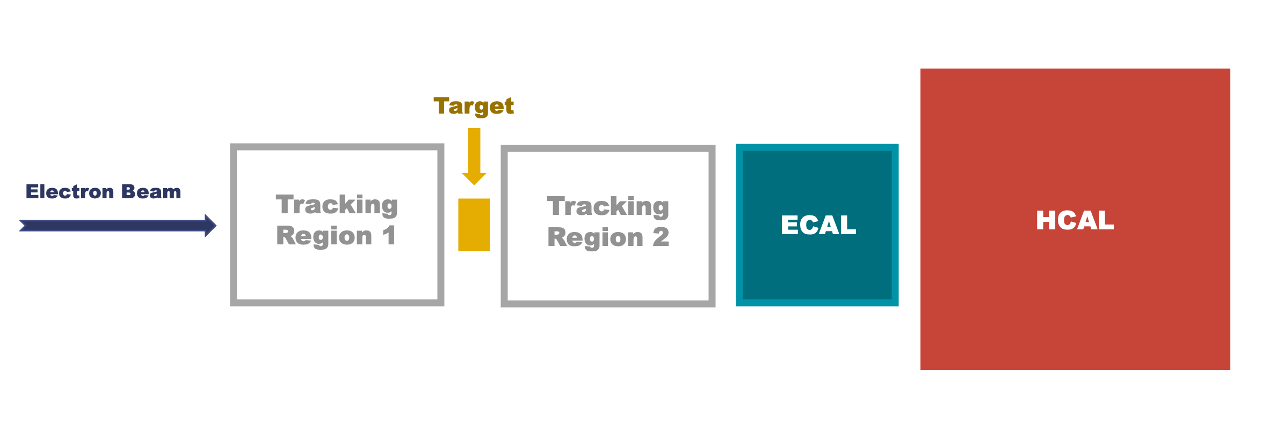}
    \caption{The detector scheme for the beam-dump experiments in this study. There are two tracking regions for reconstructing track of the incident electron and its products after interacting with the nuclei in the target. ECAL and HCAL are behind the tracking region.}
    \label{fig:detector_setup}
  \end{figure}

 The overall setup of DarkSHINE is shown in Fig.~\ref{fig:detector_setup}. The beam is an 8 GeV electron beam with a high frequency. The whole detector is composed by several main sub-detectors systems: a tracking system, an electromagnetic calorimeter (ECAL), and a hadronic calorimeter (HCAL). 

 The tracking system is composed of silicon trackers and dipole magnets, serving distinct purposes within two separate tracking regions. The first region, known as the tagging tracker, is designated for the monitoring of incident electrons, whereas the second region, the recoil tracker, is used for tracking the recoil electrons and decay products. The module of each tracker has two silicon strips placed at a small angle for more precise position measurements of each energy hit. The tungsten target with $0.1X_0$ is placed between two tracking regions. The ECAL is designed to absorb the full energy of incident particles with good energy resolution, and the HCAL is placed after ECAL, serving for detecting hadronic backgrounds and capturing muons. 
  
 \subsection{Event Simulation}
  We simulate all events with the detector setup mentioned above, utilizing specific software tools. The dark photon process is modeled using CalcHEP v3.8.10~\cite{calchep} and executed via GEANT4 v10.6.10~\cite{Geant4}. For maintaining a good modeling of the dark photon process, we directly incorporate the particles' four-momentum truth information from the generator during the dark photon process simulation. 
  Standard Model processes such as photon-nuclear interactions, electron-nuclear interactions, and photon decays into muon pairs, are simulated by GEANT4. 
  

\section{Analysis Strategy and Tracking Reconstruction}

The main background of dark photon visible decay is the QED radiative trident reaction, as depicted in Fig.\ref{fig:feynman_bkg}. While several kinematic distinctions, such as the energy and exit angle of the recoil electron, exist, the key signature of visible decay is the displaced vertex~\cite{new_fix_target}. As presented in Table \ref{tab:cut_result}, employing typical kinematics selections: (1) No. of track = 3, (2) $p(e^+)>2\ {\rm GeV}$, (3) $p({\rm hard}\ e^-)<6\ {\rm GeV}$, (4) $\theta({\rm hard}\ e^-) > 0.1\degree$, the background rejection power can reach 0.2\%, which is insufficient considering that the expected number of electrons on target reaches $3\times10^{14}$ in DarkSHINE. Therefore,  tracking and vertex reconstruction becomes vital in the quest for visible decay searching.

\begin{table}
\centering
\begin{tabular}{l c c c c c c}
    \hline
    \hline 
    Event type & total event count & cut1 & cut2 & cut3 & cut4 & efficiency \\
    \hline 
    Visible decay & 9918 & 5963 & 3889 & 3889 & 3436 & \bf{34.6\%} \\
    Inclusive background & 1.84e7 & 3.33e5 & 5.18e4 & 5.18e4 & 3.42e4 & \bf{0.19\%} \\
    \hline
\end{tabular}
\caption{The event count of the signal and inclusive background samples after kinematics selections.}
\label{tab:cut_result}
\end{table}

\subsection{GNN-based Tracking Reconstruction}
\label{sec:GNN}

We propose a novel approach to tracking reconstruction in high-energy physics, 
leveraging machine learning principles, particularly GNN. 

\subsubsection{Network Structure}

We construct 2 GNN models based on transformer convolutions to process graph-structured data using PyTorch~\cite{pyTorch} and PyG~\cite{PyG}:
\begin{itemize}
    \item \textbf{LinkNet} \textit{edge classification task for track finding}: 
    The LinkNet model is to predict if an edge is a true particle trajectory passing through the two connected nodes. 
    It employs a multi-layer perceptron (MLP) for both node and edge feature embedding, with each MLP consisting of three layers. The input dimensions are set to 6 for nodes (corresponding to spatial coordinates and magnetic field components: x, y, z, Bx, By, Bz) and 3 for edges (representing the relative polar coordinates: r, theta, phi). The hidden dimension across the network is uniformly set to 128.
    Key to LinkNet’s design is the use of Transformer Convolutional layers, which combine the strengths of CNN and transformer models. The network comprises six iterations, each with four attention heads. The node features are updated during iteration using the Transformer Convolutional layer, and the corresponding edge features are updated by adding the new features on the surrounding nodes of each edge. 
    Layer Normalization and shortcut connections are integrated into each iteration.

    For training, LinkNet is optimized using the Adam algorithm, motivated by its adaptive learning rate capabilities. The initial learning rate is set at 0.00075, with a weight decay of 1.e-5. A tiered learning rate decay schedule is implemented, reducing the learning rate by factors of 0.5, 0.2, and 0.1 at epochs 30, 40, and 50, respectively. The model undergoes training for a total of 55 epochs, a duration determined to be sufficient for convergence based on preliminary experiments.
    The loss function used is the binary cross-entropy with logits, suitable for the binary classification task intrinsic to track reconstruction — determining whether a given pair of hits belong to the same track. The batch size is set to 64.

    \item \textbf{MomNet} \textit{edge regression task for track fitting}:
    This network is specialized in predicting the momentum associated with each edge, which represents a potential particle trajectory in the detector.
    The MomNet architecture incorporates an MLP for embedding features of both nodes and edges. Each MLP contains three layers, catering to the specific dimensions of the input features. For nodes, the input dimension is 6, accounting for spatial coordinates and the magnetic field components (x, y, z, Bx, By, Bz). For edges, the input dimension is set at 4, representing the relative polar coordinates (r, theta, phi) and a link score from LinkNet that quantifies the likelihood of two nodes being connected in a physical particle track.
    MomNet also uses Transformer Convolutional layers, which are adept at handling the complex relational dynamics of the particle tracks. The network comprises 10 iterations, each with a significantly increased number of attention heads, set at 32. Each iteration of the network updates the node features using the Transformer Convolutional layer, while the corresponding edge features are refined by integrating the updated node features surrounding each edge.
    MomNet is further enhanced with layer normalization and shortcut connections in each iteration.

    For the training process, MomNet utilizes the Adam optimizer, selected for its adaptive learning rate properties. The initial learning rate is set at 0.00025, with a weight decay of 1.e-5. The learning rate undergoes a tiered decay, reducing by factors of 0.5 and 0.1 at epochs 40 and 50, respectively. The total training duration is extended to 60 epochs, slightly longer than that of LinkNet, to ensure the model's convergence given its more complex task. The batch size is maintained at 64, consistent with the LinkNet configuration.
    For the MomNet model, a specialized loss function, the RelativeHuberLoss, is employed, which combines the MSE and L1 loss. 
    The RelativeHuberLoss operates by first calculating the difference between the predicted values and the truth values. If the truth value is zero, which accounts for the case that this edge is false, and does not have true momentum along this edge, the absolute difference is used. In the other case for the truth edge, the relative difference is calculated. 
\end{itemize}

\subsubsection{Simulated Samples}

For the training phase, we use 5 visible decay signal samples with masses of 20, 50, 100, 200, and 500 MeV, each with $1\times10^6$ electron-on-target (EOT) events. In these samples, single-track events are predominant in the tagging region and 3-track events in the recoil region. The input graph is built with digitized simulated hits, derived from a mean-shift algorithm-based clustering of original simulated hits. We utilize an orthogonal set of visible decay signal samples for performance evaluation, each with $1\times10^6$ EOT.

\subsubsection{Single-track Performance Evaluation}
\begin{figure}[!h]
\begin{center}
\includegraphics[width=.45\textwidth]{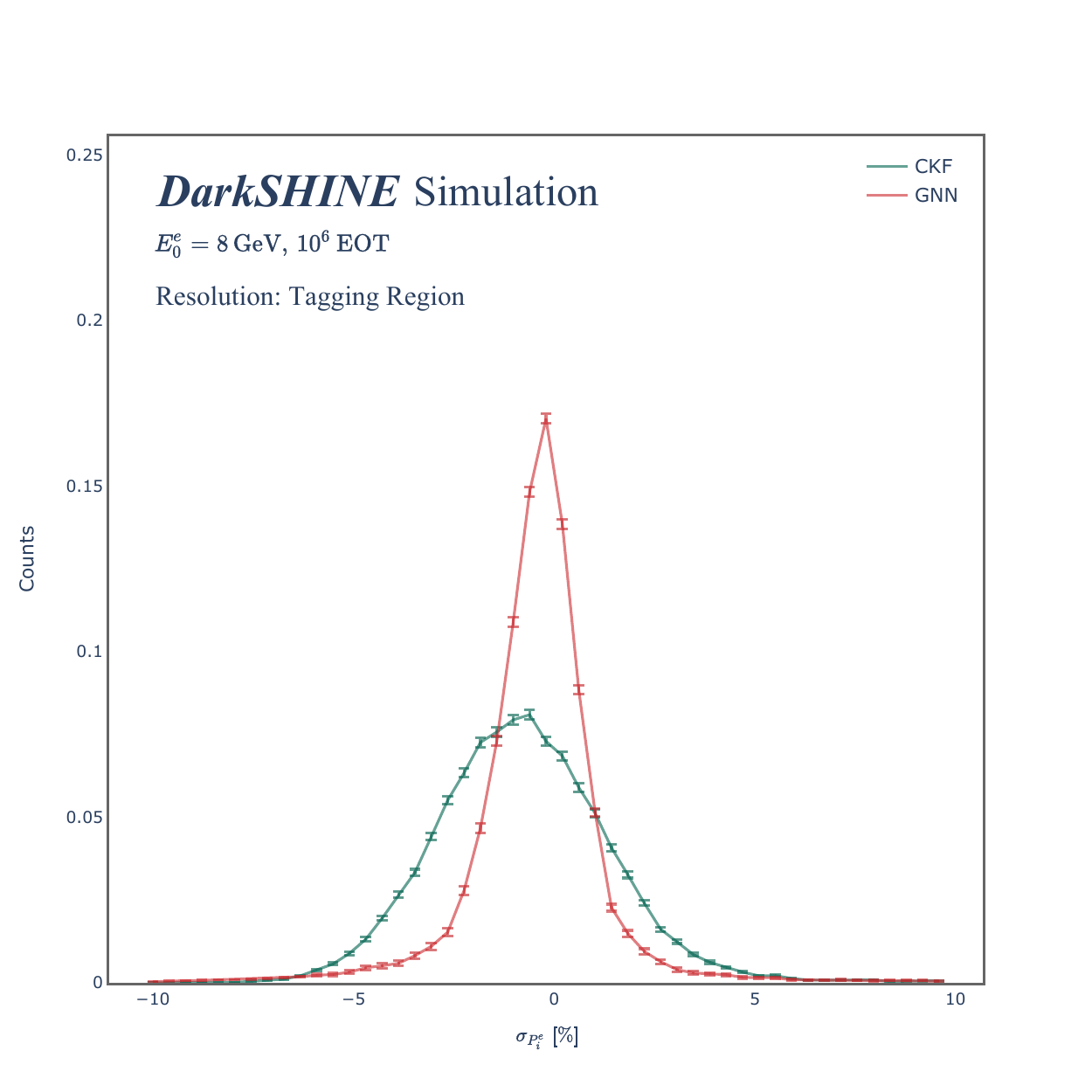}
\includegraphics[width=.45\textwidth]{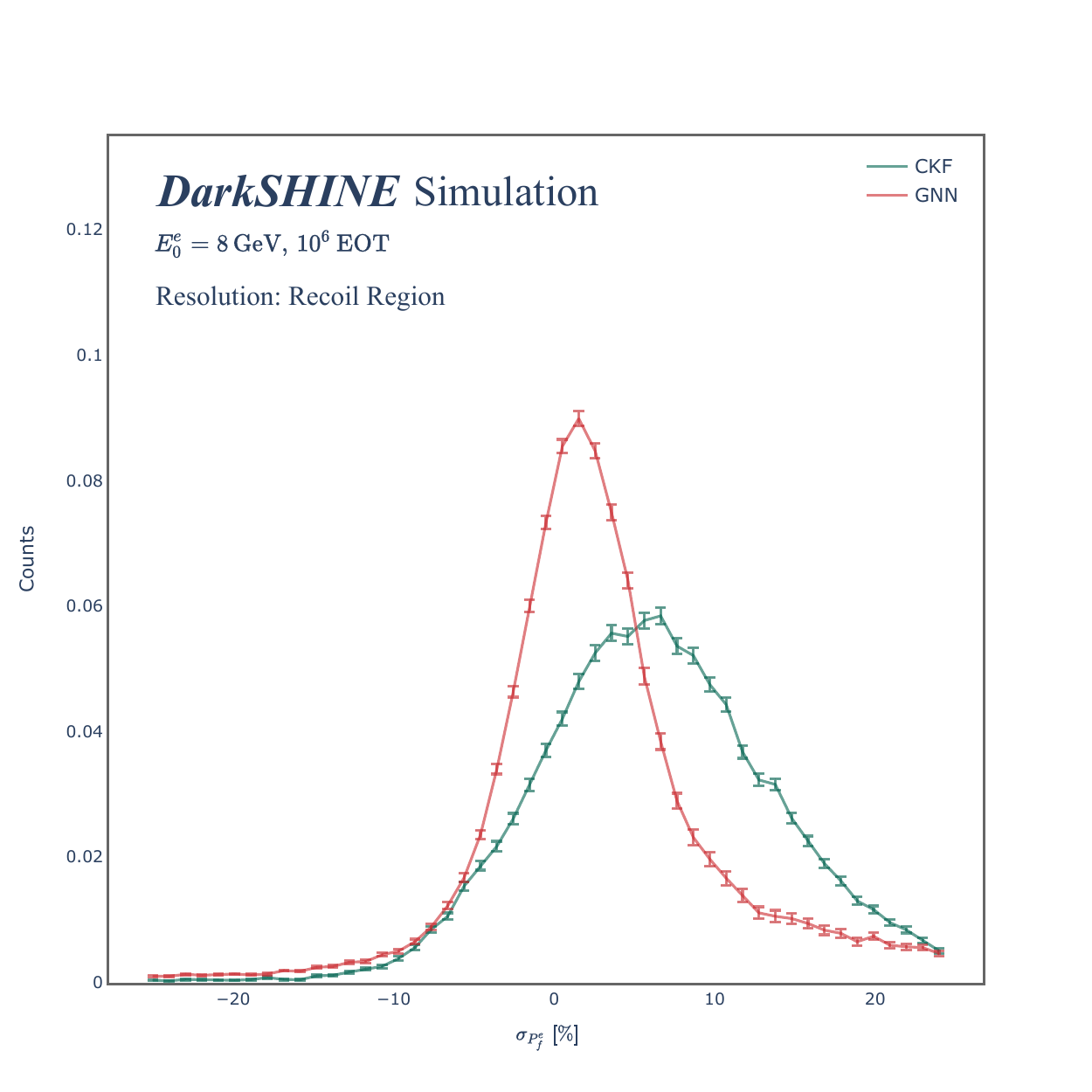}
\end{center}
\caption{
The relative uncertainty of the reconstructed momentum in the tagging region (left) and recoil region (right).
\label{fig:gnn_reso}}
\end{figure}

We evaluate the tracking performance by studying both single-track events (simple case) and 3-track events (more realistic case). The benchmark parameters are event reconstruction efficiency and track reconstruction resolution. The event reconstruction efficiency is defined as the percentage of the track events that are found and reconstructed by the reconstruction method. Track reconstruction resolution is defined as the full width at half maximum of the relative uncertainty distribution for the reconstructed track momentum. Owing to the variations in magnetic fields and tracker structures, we can assess the resolution of the reconstructed momentum in two distinct regions, as shown in Figure~\ref{fig:gnn_reso}. We have compared the GNN performance with the traditional tracking reconstruction method based on the Combined Kalman Filter (CKF) method~\cite{darkshine_exp,ckf} using an orthogonal evaluation dataset with 50000 events. 
The GNN approach can reconstruct 98.9\% single-track events, while the traditional approach gives 96.9\%. 
The comparison is shown in Table~\ref{tab:gnn_result}. GNN gives better reconstructed momentum resolution than the CKF approach, improving the resolution from 1.5\% to 0.6\% in the tagging region, and 7.5\% to 5.6\% in the recoil region.

\begin{table}
\centering
\begin{tabular}{l c c c}
    \hline
    \hline 
    Inclusive sample & CKF & GNN & Truth \\
    \hline 
    Single-track Efficiency & $96.9 \%$ & $98.9 \%$ & $99.9 \%$ \\
    Resolution (Tagging) & $1.5 \%$ & $0.6 \%$ & - \\
    Resolution (Recoil) & $7.5 \%$ & $5.6 \%$ & - \\
    \hline
\end{tabular}
\caption{The reconstruction efficiency and momentum resolution comparison between the CKF and GNN methods.}
\label{tab:gnn_result}
\end{table}

\begin{table}
\centering
\begin{tabular}{l c c}
\hline
\hline Average Time [sec] & CKF & GNN \\
\hline No pile-up & 0.179 & 0.002 \\
\hline Pile-up $<\mu \approx 10>$ & $300$ & 0.010 \\
\hline
\end{tabular}
\caption{Comparative analysis of average computational time per event (in seconds) for the CKF method and the GNN method with and without pile-up.}
\label{tab:compute_time}
\end{table}

\subsubsection{Multi-track Performance Evaluation}
\begin{figure}[!h]
\begin{center}
\includegraphics[width=.31\textwidth]{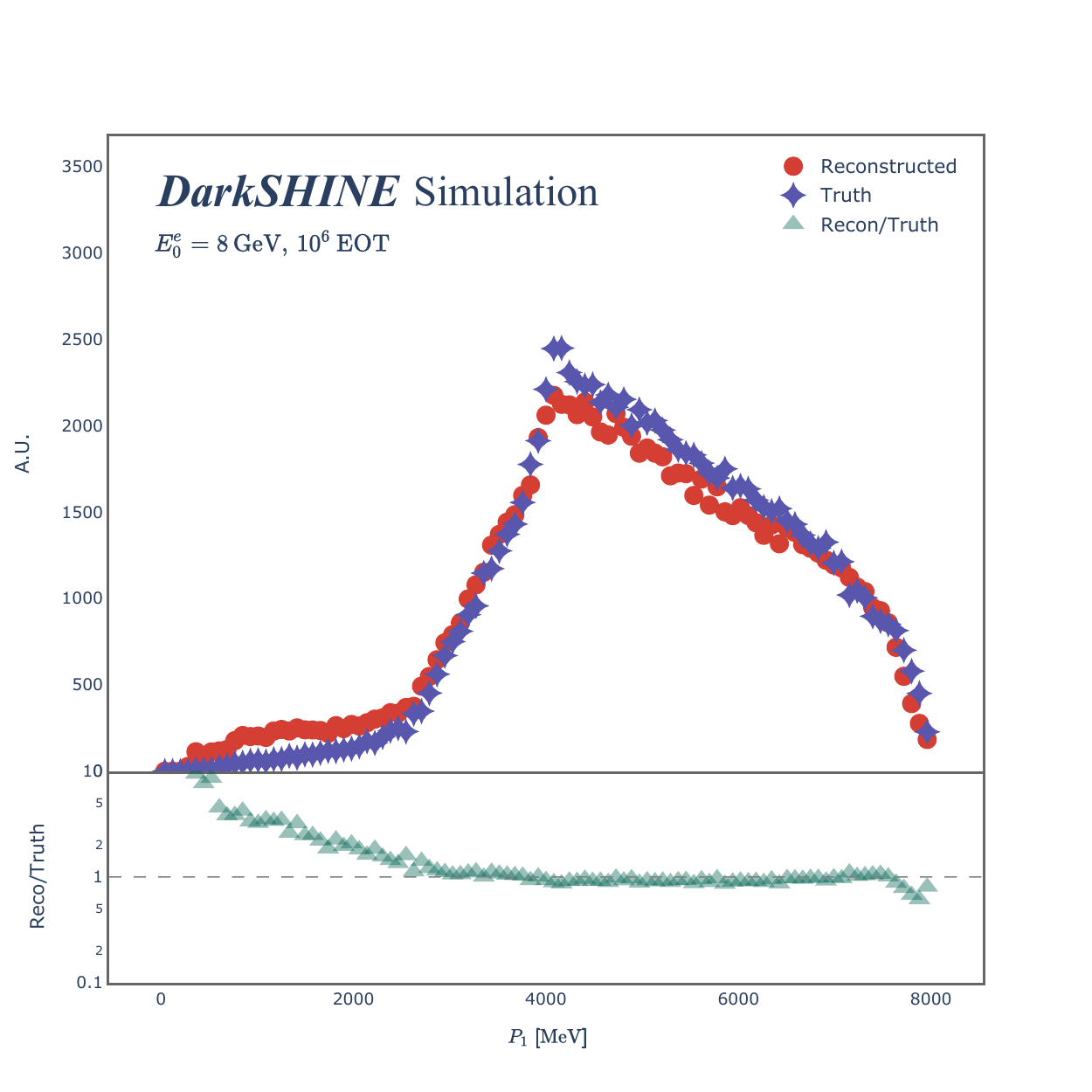}
\includegraphics[width=.31\textwidth]{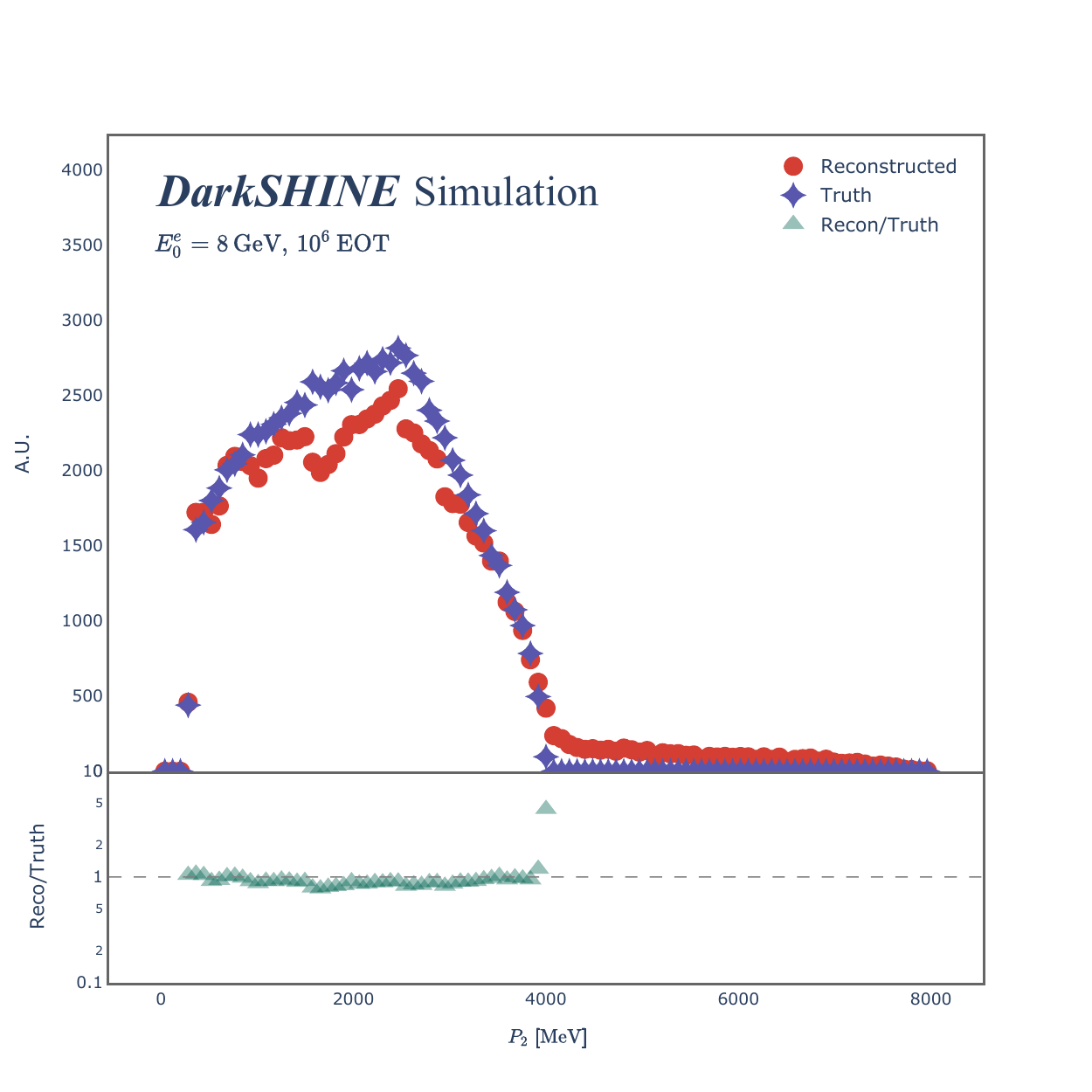}
\includegraphics[width=.31\textwidth]{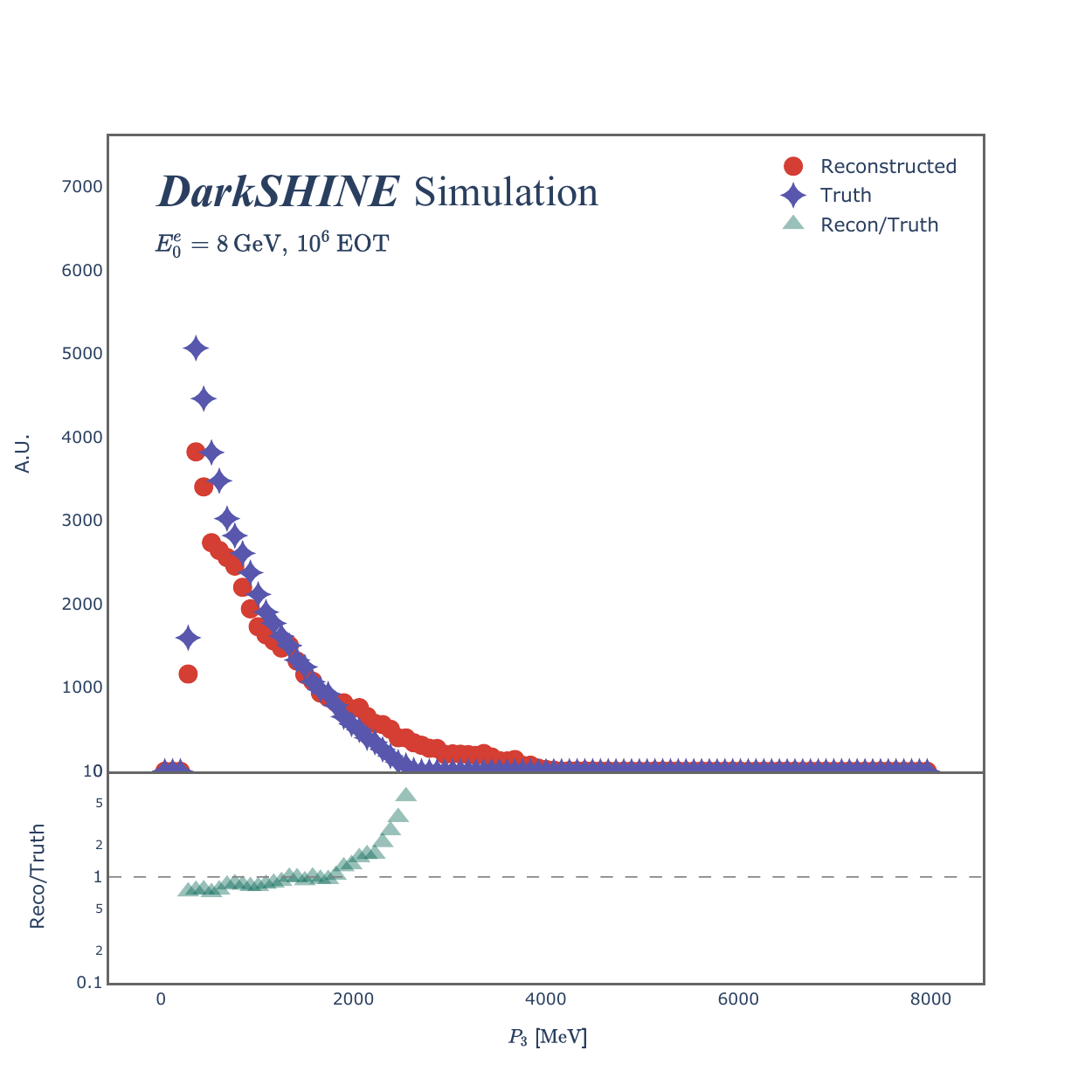}
\end{center}
\caption{
The momentum distribution for reconstruction and truth for the leading track (left), subleading track (middle), and subsubleading track (right).
\label{fig:gnn_mom}}
\end{figure}

We evaluate the GNN performance to find 3-track events in the visible decay mode. The 3-track reconstruction efficiencies for 5 mass points are listed in Table~\ref{tab:gnn_3-track}. The low mass (long lifetime) region is the most sensitive region for beam dump experiments. GNN outperforms the CKF method by 30\% to 88\% in the region of interest ($< 100$~MeV) and gives overall reconstruction efficiencies around 60\%. The reconstructed momentum for each track is also shown in Figure~\ref{fig:gnn_mom}. There is a selection criterion requiring $P_{i} \ge 150$~MeV to ensure a good track quality in the tracker region.
The application of GNN significantly improves the computational efficiency in terms of time per event, compared to the traditional CKF method, which is summarized in Table~\ref{tab:compute_time}. 

\begin{table}[]
    \centering
    \begin{tabular}{c c c c c c}
    \hline
    \hline
    \(m_{A^{\prime}}[\mathrm{MeV}]\) & \(\mathbf{20}\) & \(\mathbf{50}\) & \(\mathbf{100}\) & \(\mathbf{200}\) & \(\mathbf{500}\) \\
    \hline 
    CKF & \(31.8 \%\) & \(43.7 \%\) & \(51.5 \%\) & \(61.2 \%\) & \(68.4 \%\) \\
     
    GNN & \(59.9 \%\) & \(61.9 \%\) & \(66.8 \%\) & \(65.2 \%\) & \(50.3 \%\) \\
    \hline
    \end{tabular}
    \caption{The 3-track reconstruction efficiency with respect to truth 3-track events.}
    \label{tab:gnn_3-track}
\end{table}

\subsubsection{Future Prospects}

The current study has demonstrated the effectiveness of GNNs for track fitting and finding tasks. These networks effectively leverage the natural graph structure inherent in detector data, creating a powerful framework for these complex tasks. 
Further research into vertex finding is also a promising direction. Several deep-learning methods have been proposed and used for vertex finding based on various experimental setups and constraints. Exploring more architectures and approaches becomes crucial. For instance, the use of Point Cloud Networks or Deep Sets could offer innovative strategies for addressing vertex-finding tasks, taking advantage of their unique capabilities for dealing with point-like structures or variable-sized sets of data.

\subsection{Dark Photon Search in Visible Decay Mode}

\begin{figure}[htb]
\centering
\begin{subfigure}{0.45\textwidth}
\includegraphics[width=\textwidth]{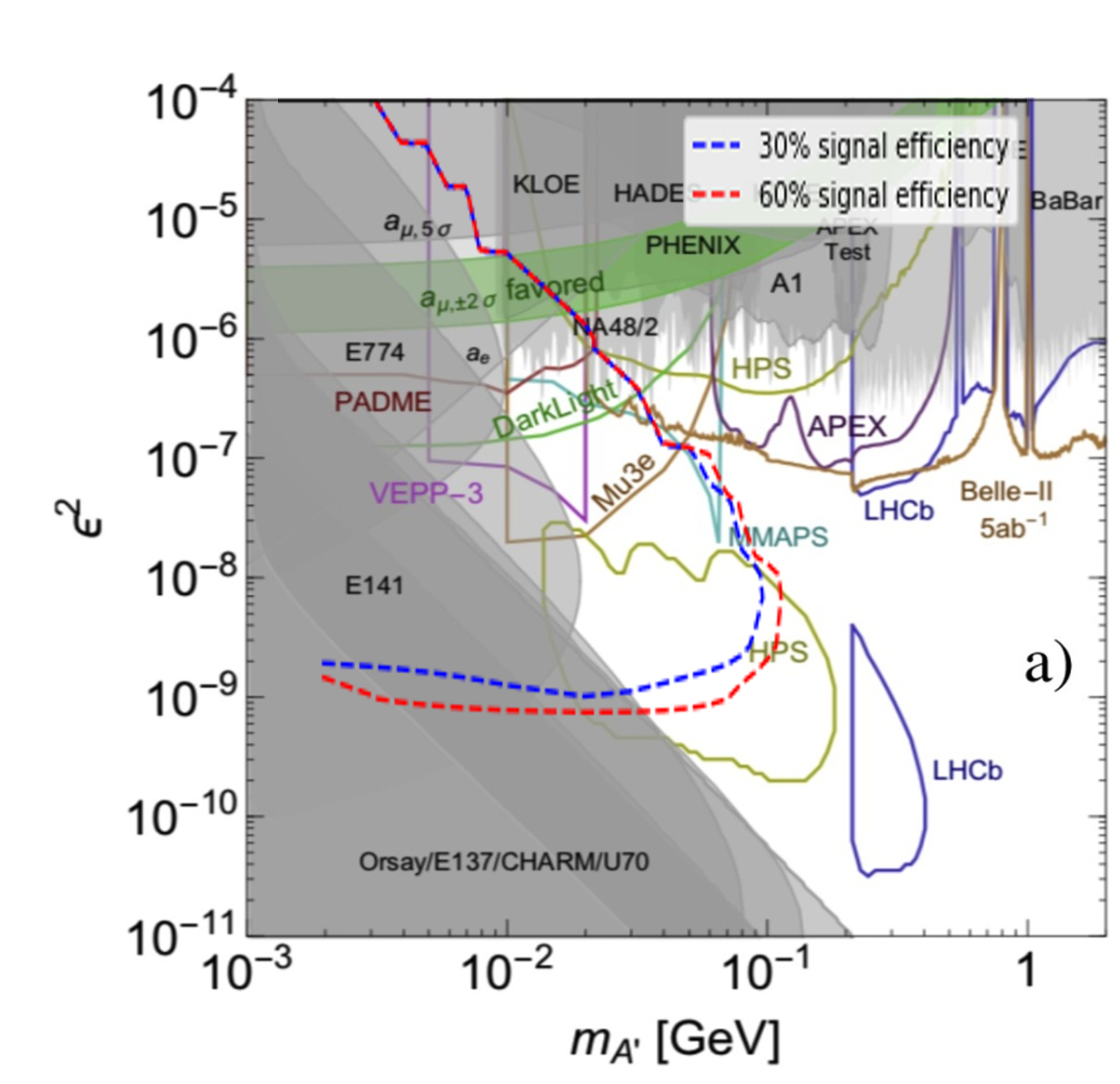}
\caption{Signal Efficiency Comparison}
\label{fig:signal_region_a}
\end{subfigure}
\begin{subfigure}{0.45\textwidth}
\includegraphics[width=\textwidth]{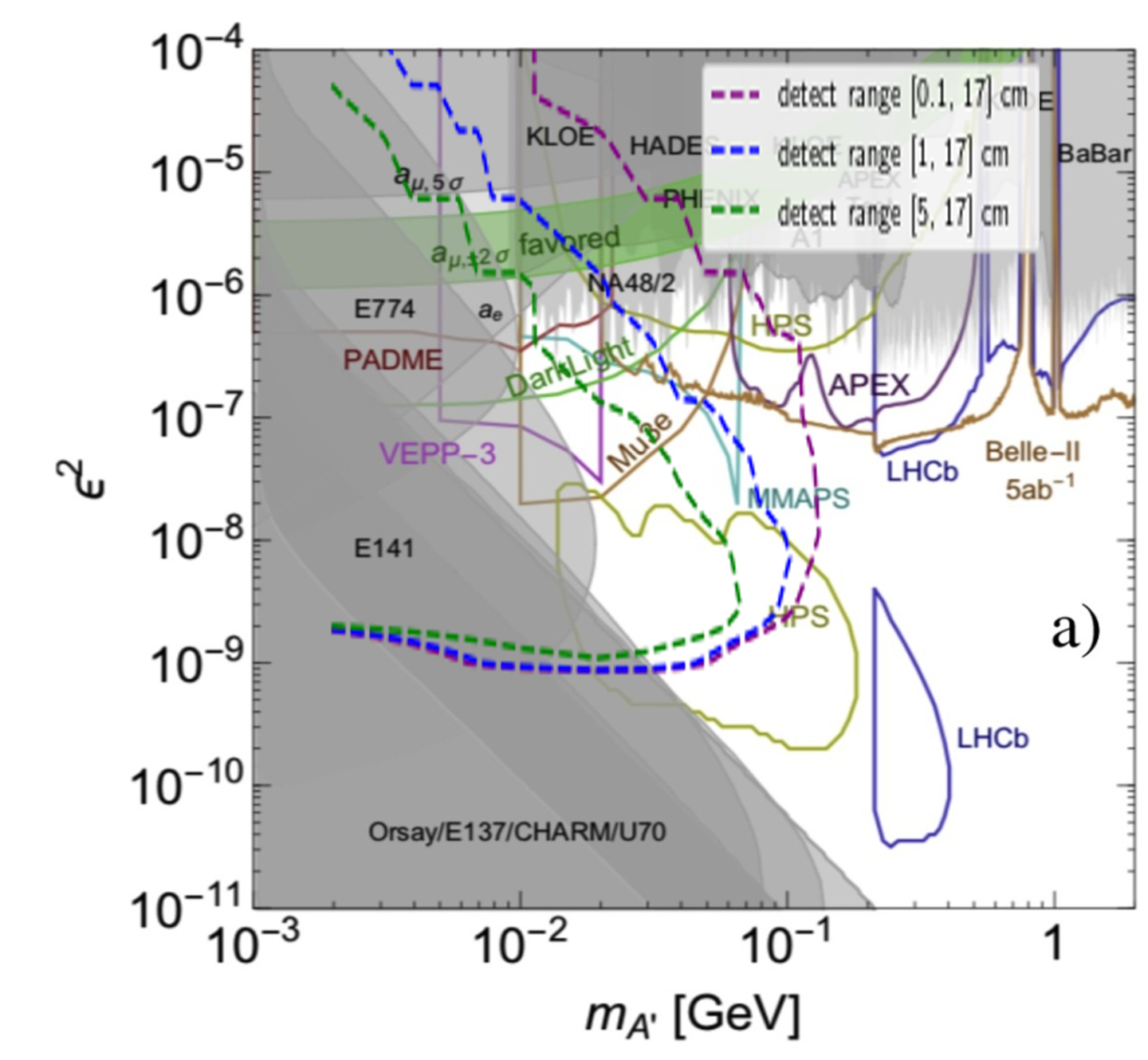}
\caption{Vertex Resolution Comparison}
\label{fig:signal_region_b}
\end{subfigure}
\caption{90\% confidence level exclusion
region for dark photon search in the visible decay mode for the DarkSHINE project when comparing different conditions: (a) signal efficiency comparison (b) vertex resolution comparison. Gray indicates regions that have been explored by other experiments~\cite{searching}.}
\label{fig:signal_region}
\end{figure}

With the above improvements in both tracking and vertex reconstruction, the residual background level is expected to be 10 out of $3\times10^{14}$ EOT. The 90\% confidence level exclusion region for DarkSHINE experiment in the search for visible decays is calculated in Fig.~\ref{fig:signal_region}. According to Sec.~\ref{sec:GNN}, the improved signal reconstruction efficiency for GNN method is expected to be around 60\% in the most sensitive mass region (less than 100~MeV), while the efficiency for CKF method is between 30\% and 50\%. We have compared the two cases of signal efficiency being 30\% and 60\%, the result is shown in Fig.~\ref{fig:signal_region_a}. Furthermore, with improved track momentum resolution, GNN method is expected to provide better vertex reconstruction resolution, which is important for signal searching in the visible decay mode. We present three projected signal exclusion limits when the minimal vertex detection distance is set to be 0.1~mm, 1~mm, and 5~mm respectively, as shown in Fig.~\ref{fig:signal_region_b}. Taking $\epsilon^2=10^{-8}$ as an example, the upper limit on thr dark photon mass $m_A\prime$ can be extended to 55 MeV, 95 MeV and 155 MeV with vertex resolutions of 5 mm, 1 mm and 0.1 mm respectively. 

\section{Conclusion and Outlook}

We have demonstrated that
in the single track and 3-tracks scenario, GNN approach outperforms the traditional
approach based on Combinatorial Kalman Filter (CKF) and improves
the 3-track event reconstruction efficiency by 88\% in the most sensitive signal region. Furthermore, we have showed that improving the minimal vertex detection distance has significant impact on the signal sensitivity in dark photon
searches in the visible decay mode. The exclusion upper limit on the dark
photon mass $m_A\prime$ can be improved by up to a factor of 3 by
reducing the minimal vertex distance from 5 mm to 0.1 mm.


\subsubsection{Acknowledgements}
This work was supported by the National Natural Science Foundation of China (Grant No. 12211540001). 
We thank for the support from from Key Laboratory for Particle Astrophysics and Cosmology (KLPPAC-MoE), Shanghai Key Laboratory for Particle Physics and Cosmology (SKLPPC).

%
%
%
%

\end{document}